\documentclass[12pt]{article}
\usepackage{amssymb}

\def\hybrid{\topmargin 0pt      \oddsidemargin 0pt
        \headheight 0pt \headsep 0pt
        \voffset=-0.5cm
        \hoffset=-0.25in
        \textwidth 6.75in
        \textheight 9.5in       
        \marginparwidth 0.0in
        \parskip 5pt plus 1pt   \jot = 1.5ex}
\catcode`\@=11
\def\marginnote#1{}

\newcount\hour
\newcount\minute
\newtoks\amorpm
\hour=\time\divide\hour by60 \minute=\time{\multiply\hour by60
\global\advance\minute by-\hour}
\edef\standardtime{{\ifnum\hour<12 \global\amorpm={am}%
        \else\global\amorpm={pm}\advance\hour by-12 \fi
        \ifnum\hour=0 \hour=12 \fi
        \number\hour:\ifnum\minute<10 0\fi\number\minute\the\amorpm}}
\edef\militarytime{\number\hour:\ifnum\minute<10 0\fi\number\minute}

\def\draftlabel#1{{\@bsphack\if@filesw {\let\thepage\relax
   \xdef\@gtempa{\write\@auxout{\string
      \newlabel{#1}{{\@currentlabel}{\thepage}}}}}\@gtempa
   \if@nobreak \ifvmode\nobreak\fi\fi\fi\@esphack}
        \gdef\@eqnlabel{#1}}
\def\@eqnlabel{}
\def\@vacuum{}
\def\draftmarginnote#1{\marginpar{\raggedright\scriptsize\tt#1}}
\def\draftlabel#1{{\@bsphack\if@filesw {\let\thepage\relax
   \xdef\@gtempa{\write\@auxout{\string
      \newlabel{#1}{{\@currentlabel}{\thepage}}}}}\@gtempa
   \if@nobreak \ifvmode\nobreak\fi\fi\fi\@esphack}
        \gdef\@eqnlabel{#1}}
\def\@eqnlabel{}
\def\@vacuum{}
\def\draftmarginnote#1{\marginpar{\raggedright\scriptsize\tt#1}}

\def\draft{\oddsidemargin -.5truein
        \def\@oddfoot{\sl preliminary draft \hfil
        \rm\thepage\hfil\sl\today\quad\militarytime}
        \let\@evenfoot\@oddfoot \overfullrule 3pt
        \let\label=\draftlabel
        \let\marginnote=\draftmarginnote
   \def\@eqnnum{(\theequation)\rlap{\kern\marginparsep\tt\@eqnlabel}%
\global\let\@eqnlabel\@vacuum}  }

\def\numberbysection{\@addtoreset{equation}{section}
        \def\theequation{\thesection.\arabic{equation}}}

\def\underline#1{\relax\ifmmode\@@underline#1\else
        $\@@underline{\hbox{#1}}$\relax\fi}

\def\titlepage{\@restonecolfalse\if@twocolumn\@restonecoltrue\onecolumn
     \else \newpage \fi \thispagestyle{empty}\c@page\z@
        \def\thefootnote{\fnsymbol{footnote}} }

\def\endtitlepage{\if@restonecol\twocolumn \else  \fi
        \def\thefootnote{\arabic{footnote}}
        \setcounter{footnote}{0}}  

\numberbysection \hybrid

\newcounter{mo}

\newcommand{\tr}{{\rm tr}}
\newcommand{\ti}[1]{\tilde{#1}}

\newcommand{\la}{\lambda}

\newcommand{\al}{\alpha}
\newcommand{\be}{\beta}
\newcommand{\ga}{\gamma}
\newcommand{\om}{\omega}

\newcommand{\MatM}{ {\rm Mat}(M,\mathbb C) }

\newcommand{\mC}{\mathbb C}

\def\beq{\begin{equation}}
\def\eq{\end{equation}}
\def\p{\partial}

\newcommand{\mats}[4]{\left(\begin{array}{cc}{#1}&{#2}\\ {#3}&{#4}
\end{array}\right)}

\begin{document}

\setcounter{page}{1}

\date{}
\date{}
\vspace{30mm}

\begin{flushright}
 ITEP-TH-35/19\\
\end{flushright}
\vspace{0mm}

\begin{center}
\vspace{0mm}
 {\LARGE{Quantum-classical duality for Gaudin magnets}}
\\ \vspace{4mm} {\LARGE{with boundary }}
\\
\vspace{15mm} {\large \ \ {M. Vasilyev}\,{\small $^{\flat\,
 \diamondsuit\,\ddagger}$}
 \ \ \ \ \ {A. Zabrodin}\,{\small $^{\flat\, \S\, \sharp }$}
 \ \ \ \ \ {A. Zotov}\,{\small $^{\diamondsuit\, \ddagger\,
\natural}$} }
 \vspace{10mm}

 \vspace{1mm} $^\flat$ --
 {\small{\rm Skolkovo Institute of Science and Technology,
 Nobel str. 1, Moscow,  143026 Russia}}
\\
\vspace{1mm} $^\diamondsuit$ -- {\small{\rm
 Steklov Mathematical Institute of Russian Academy of Sciences,\\ Gubkina str. 8, Moscow,
119991,  Russia}}
 \\
 \vspace{1mm}$^\S$ - {\small{\rm National Research University Higher School of Economics,  \\
  Myasnitskaya str. 20,
Moscow, 101000, Russia}}
  \\
 \vspace{1mm} $^\sharp$ --
 {\small{\rm Institute of
Biochemical Physics, Kosygina str. 4, 119334, Moscow, Russia}}
\\
%
%
 \vspace{1mm} $^\ddagger$ -- {\small{\rm Institute for Theoretical and
 Experimental
 Physics of
  NRC ''Kurchatov Institute'',\\
 B. Cheremushkinskaya str. 25,  Moscow, 117218, Russia}}
\\
  \vspace{1mm} $^\natural$ -- {\small{\rm Moscow Institute of Physics and Technology,\\
  Institutsky lane 9, Dolgoprudny,
 Moscow region, 141700, Russia}}

\end{center}

\begin{center}\footnotesize{{\rm E-mails:}{\rm\
 mikhail.vasilyev@phystech.edu,\ zabrodin@itep.ru,\ zotov@mi-ras.ru}}\end{center}
%
%

 \begin{abstract}
 We establish a remarkable relationship between the quantum Gaudin models with
 boundary and the classical many-body integrable systems of Calogero-Moser type
 associated with the root systems of classical Lie algebras (B, C and D).
 We show that under identification of spectra of the Gaudin
 Hamiltonians $H_j^{\rm G}$ with
 particles velocities $\dot q_j$ of the classical model
all integrals of motion of the latter take zero values.
 This is the generalization of the quantum-classical duality observed earlier for
 Gaudin models with periodic boundary conditions and Calogero-Moser
 models associated with the root system of the type A.

 \end{abstract}

\bigskip



\section{Introduction: quantum-classical duality}
 \setcounter{section}{1}
 \setcounter{equation}{0}

 The  quantum-classical duality for integrable systems is a certain relation
 between the spectrum of an inhomogeneous quantum spin chain
 (or its limit to a model of the Gaudin type)
 and intersection of two
 Lagrangian submanifolds in the $2N$-dimensional phase space of a classical relativistic
$N$-body integrable system of the Ruijsenaars-Schneider type
 (or its non-relativistic limit which is the Calogero-Moser system). Namely, the
 first
 Lagrangian manifold is the $N$-dimensional
 hyperplane corresponding to fixing all coordinates $q_j$ of
 the classical particles, and the second one is the level set
 of the $N$ integrals of motion in involution. Since their dimensions are
 complimentary, they intersect in a finite number of points.
 The essence of the quantum-classical
 duality is that the values of the particles velocities $\dot q_j$ at the
 intersection points provide spectra of the quantum Hamiltonians of the inhomogeneous
 spin chain (or the Gaudin model). Different intersection points correspond to
 different eigenstates of the commuting quantum Hamiltonians.

 The relation of this type was independently observed several times
 from several different viewpoints and using different
 approaches including the ones related to quantum cohomologies
 \cite{Givental}, the Knizh\-nik-Za\-mo\-lod\-chi\-kov equations \cite{MTV},
 the (modified) Kadom\-tsev-Petviashvili hierarchy \cite{Zabrodin1} and the quiver
 gauge theories \cite{GK13}. For quantum models with rational $R$-matrices,
 the quantum-classical duality in its most general form  was formulated
 and proved in the paper \cite{GZZ}
 by means of the algebraic Bethe ansatz
 method and certain
 relations for characteristic polynomials
 of the Lax matrices of the classical integrable many-body systems.
In this paper we follow the latter approach.

 Let us consider, in more detail,
 the simplest example illustrating the duality. On the quantum side, we have
  the ${\mathfrak {gl}}_2$ Gaudin XXX magnet \cite{Gaudin}.
The quantum ${\mathfrak {gl}}_2$ rational \underline{Gaudin model}
  describes the spin-exchange type long-range interaction of $N$ spins $1/2$ on
the complex plane (more precisely, on $\mC P^1$).
 It is defined by the set of commuting
 Hamiltonians
 \beq\label{w01}
 \begin{array}{c}
  \displaystyle{
 {\bf H}_i^{\hbox{\tiny{G}}}={\bf w}^{(i)}+\hbar\sum\limits_{k\neq i}
 \frac{{\bf P}_{ik}}{z_i-z_k}\,, \quad i=1, \ldots ,N\,,
 }
 \end{array}
  \eq
 where ${\bf P}_{ik}$ is the permutation operator of two complex linear spaces
 $V_i\cong \mC ^2$ and $V_k\cong \mC ^2$
 written in either the
 standard matrix basis $E_{ab}$ ($a,b=1,2$) in ${\rm Mat}(2,\mC)$
 ($(E_{ab})_{a'b'}=\delta_{aa'}\delta_{bb'}$)
 or in the Pauli
 matrices basis $\sigma_0=\left (\begin{array}{cc}
 1&0\\0&1\end{array}\right )$,
 $\sigma_1=\left (\begin{array}{cc}
 0&1\\1&0\end{array}\right )$,
 $\sigma_2=\left (\begin{array}{cc}
 0&-i\\i&0\end{array}\right )$,
 $\sigma_3=\left (\begin{array}{cc}
 1&0\\0&-1\end{array}\right )$,
 as follows:
 \beq\label{w02}
 \begin{array}{c}
  \displaystyle{
 {\bf P}_{ik}=\sum\limits_{a,b=1}^2 E_{ab}^{(i)}E_{ba}^{(k)}=
 \frac{1}{2}\sum\limits_{\al=0}^3\sigma_\al^{(i)}\sigma_\al^{(k)}\,,
 \qquad
 E_{ab}^{(i)}=1\otimes 1 \ldots \underbrace{1\otimes E_{ab}\otimes
 1}_{\hbox{\tiny{at the
 i-th place}}} \ldots 1\otimes 1\,.
 }
 \end{array}
  \eq
 It acts non-trivially on the $i$-th and $k$-th components of the
 tensor product
 $\displaystyle{{\mathcal H}=\otimes_{i=1}^{N}V_i}$, $V_i\cong\mC^2$, which is the
 Hilbert space of states of the quantum model. The term ${\bf w}^{(i)}$ in (\ref{w01})
 is a constant matrix ${\bf w}=\mats{\om}{0}{0}{-\om}$ acting
 non-trivially in the $i$-th tensor component.
 It is called the twist
 matrix and $\omega$ is called the twist parameter. Due to commutativity of the Gaudin
 Hamiltonians, $[{\bf H}_i^{\hbox{\tiny{G}}},{\bf H}_j^{\hbox{\tiny{G}}}]=0$,
 the eigenvalue problems
 \beq\label{w03}
 \begin{array}{c}
  \displaystyle{
 {\bf H}_i^{\hbox{\tiny{G}}}\psi=H_i^{\hbox{\tiny{G}}}\psi\,,
 \quad \psi\in{\mathcal H}\,,\quad i=1, \ldots ,N
 }
 \end{array}
  \eq
have a complete set of common solutions. The algebraic Bethe ansatz provides the
following answer for the spectrum $H_i^{\hbox{\tiny{G}}}$ of the
Hamiltonians (\ref{w01}):
 \beq\label{w04}
\begin{array}{c}
  \displaystyle{
H_i^{\hbox{\tiny{G}}}= \om+\sum\limits_{k\neq
i}^N\frac{\hbar}{z_i-z_k}+
\sum\limits_{\ga=1}^{M}\frac{\hbar}{\mu_\ga-z_i}\,,\quad
i=1, \ldots ,N\,,
 }
\end{array}
  \eq
 where the set of Bethe roots $\{\mu_\al\,,\al=1,...,M\}$ is a
 solution of
 the system of $M$ Bethe equations (BE):
 \beq\label{w05}
\begin{array}{c}
  \displaystyle{
2\om+ \hbar\sum\limits_{k=1}^N\frac{1}{\mu_\al-z_k}
=2\hbar\sum\limits_{\ga\neq\al}^{M}
\frac{1}{\mu_\al-\mu_\ga}\,,\quad \al=1, \ldots ,M\,.
 }
\end{array}
  \eq
The integer parameter $M\leq [N/2]$ is the number of the overturned spins
in the eigenstate $\psi$. For
example, if the vacuum vector is chosen to consist of all spins looking up,
then $M$ is the number of spins looking down in the state vector $\psi$.
More precisely, since the Gaudin Hamiltonians commute with the operator
$${\bf M}=\sum_{i=1}^N E_{22}^{(i)}=\frac{1}{2}\sum_{i=1}^N (\sigma_0^{(i)}-\sigma_3^{(i)}),$$
the state vector $\psi$ which solves (\ref{w03}) is also
an eigenvector of the operator ${\bf M}$,
then $M$ is the corresponding eigenvalue: ${\bf M}\psi = M\psi$.

 On the classical side, we deal with the
 $N$-body rational ${\mathfrak{gl}}_N$
\underline{Calogero-Moser model} \cite{Calogero} (associated with the root
system ${\rm A}_{N-1}$). It is described by the Hamiltonian
 \beq\label{w07}
 \begin{array}{c}
   \displaystyle{
 H^{\hbox{\tiny{CM}}}= \frac{1}{2}\sum\limits_{i=1}^N
p_i^2- \sum\limits_{i<j}^N \frac{g^2}{(q_i-q_j)^2}
 }
 \end{array}
  \eq
  with the canonical Poisson brackets between $p_i$ and $q_j$, where
 $g \in\mC$ is the coupling constant.
The equations of motion are $\dot q_i =p_i$ and
 \beq\label{w08}
 \begin{array}{c}
   \displaystyle{
\dot p_i=\ddot q_i =-\sum_{k\neq i}\frac{2g^2}{(q_i-q_k)^3}\,,
\quad i=1\,, \ldots \,,N\,.
 }
 \end{array}
  \eq
  The model is known to be integrable and possessing Lax representation
  $\dot L^{\hbox{\tiny{CM}}}=[L^{\hbox{\tiny{CM}}}, M^{\hbox{\tiny{CM}}}]$.
The Lax matrix $L^{\hbox{\tiny{CM}}}$ of size $N\! \times \! N$ is \cite{OP}
 \beq\label{w09}
 \begin{array}{c}
   \displaystyle{
L^{\hbox{\tiny{CM}}}_{ij}(\{\dot q_j\}\,,\{q_j\}\,, g)=
\delta_{ij}{\dot q}_i+g \frac{1-\delta_{ij}}{q_i-q_j}\,,\ \ \
i,j=1\,, \ldots \,,N\,.
 }
 \end{array}
  \eq
Its eigenvalues $\hbox{Spec}(L^{\hbox{\tiny{CM}}})=(I_1, \ldots ,I_N)$
are integrals of motion. The higher Hamiltonians in involution are given by
$\displaystyle{H^{\hbox{\tiny{CM}}}_k =\frac{1}{k}\, \tr
\left(L^{\hbox{\tiny{CM}}}\right)^k =\frac{1}{k}\sum_{i=1}^N I_i^k}$. In particular,
$H^{\hbox{\tiny{CM}}}=H^{\hbox{\tiny{CM}}}_2$. The $M$-matrix is given by
 \beq\label{w091}
 \begin{array}{c}
   \displaystyle{
M^{\hbox{\tiny{CM}}}_{ij}= \delta_{ij}\sum\limits_{k\neq
i}\frac{g}{(q_i-q_k)^2}-(1-\delta_{ij})
\frac{g}{(q_i-q_j)^2}\,,\ \ \ i,j=1\,, \ldots \,,N.
 }
 \end{array}
  \eq

The statement of the \underline{quantum-classical duality} is as
follows. Given the Lax matrix (\ref{w09}),
consider the substitution
 \beq\label{w10}
    \displaystyle{
 q_j=z_j\,,\qquad g = \hbar \qquad \mbox{and} \qquad
\dot q_j=
H_j^{\hbox{\tiny{G}}}\,, \ \
j=1\,,...\,,N\,,
 }
 \eq
where $H_j^{\hbox{\tiny{G}}}$ are eigenvalues of the Gaudin
Hamiltonians given by (\ref{w04}). Then, if the
$M$ Bethe roots $\mu_\al$ satisfy the Bethe equations (\ref{w05}),
the spectrum of the Lax matrix takes
the form
 \beq\label{w11}
 \begin{array}{c}
 \hbox{Spec} \, L^{\hbox{\tiny{CM}}}
 \left ( \{ 
 H_j^{\hbox{\tiny{G}}} \},
 \left \{ z _j\right \} \,, \hbar \right )\Bigr |_{BE}
 =\big\{\underbrace{\om\,,\ldots\,,\om}_{N-M}\,, \,
 \underbrace{-\om\,,\ldots\,,-\om}_{M}\big\}\,,
\end{array}
  \eq
 that is the eigenvalues of the Lax matrix are equal to the elements of the diagonal
 twist matrix, with the multiplicities determined by the eigenvalue
 of the operator ${\bf M}$
 (the multiplicities are the numbers of spins looking up and down in the eigenstate
 $\psi$ of the Hamiltonians ${\bf H}_j^{\hbox{\tiny{G}}}$ with the eigenvalues
 $H_j^{\hbox{\tiny{G}}}$).
 In particular, in the case when the twist is absent ($\omega =0$), all eigenvalues
 of the Lax matrix are equal to zero, i.e., the matrix consists of Jordan cells.
 This means that in this case all the Hamiltonians take zero values.
 Therefore, we see that the spectrum of the Gaudin Hamiltonians
 ${\bf H}_j^{\hbox{\tiny{G}}}$ is indeed reproduced
 as the $\dot q_j =p_j$-coordinates of the intersection points of the level zero set
 of the higher Calogero-Moser Hamiltonians in involution
 (an $N$-dimensional hypersurface embedded in the phase space)
 and the $N$-dimensional hyperplane corresponding to the fixed values of
 the coordinates $q_j =z_j$.

The proof of the quantum-classical duality (i.e.,
of the statement (\ref{w11})) is technically based on a remarkable determinant
identity \cite{GZZ}. Consider the pair of matrices
 \beq\label{w13}
\begin{array}{c}
 \displaystyle{
  {\mathcal{L}}_{ij}=
\delta_{ij}\left(\om+\sum\limits^N_{k\neq
i}\frac{\hbar}{q_i-q_k}+\sum\limits_{\ga=1}^M\frac{\hbar}{\mu_\ga-q_i}
\right)+(1-\delta_{ij})\frac{\hbar}{q_i-q_j}\,,\quad i,j=1, \ldots ,N }
\end{array}
  \eq
and
 \beq\label{w14}
\begin{array}{c}
\displaystyle{{\widetilde {\mathcal{L}}}_{\al\be} =
\delta_{\al\be}\left(
\om-\sum\limits^M_{\ga\neq\al}\frac{\hbar}{\mu_\al\!-\!\mu_\ga}
-\sum\limits^N_{k=1}\frac{\hbar}{q_k\!-\!\mu_\al}\right)+
\left(1-\delta_{\al\be}\right) \frac{\hbar}{\mu_\al\!-\!\mu_\be}\,,
 \quad \al,\be=1, \ldots ,M\,.
}
\end{array}
  \eq
Then the following relation between the characteristic polynomials
of these matrices holds:
 \beq\label{w15}
  \displaystyle{
 \det_{N\times N}
 \Bigl ({\mathcal{L}}-\la I\Bigr )=(\om-\la)^{N-M}
\det_{M\times M} \Bigl ({\widetilde {\mathcal{L}}} -\la I\Bigr )\,.
 }
  \eq
(Here $I$ is the identity matrix.) The matrix ${\mathcal L}$ is the
Lax matrix (\ref{w09}) with the identification $\dot
q_i=H_i^{\hbox{\tiny{G}}}$ (where the parameters $\{\mu_\gamma \}$
entering $H_i^{\hbox{\tiny{G}}}$ are regarded as free parameters,
i.e., here they are not supposed to satisfy the Bethe equations).
Identities of this type also find their applications in computation
of the scalar products of Bethe vectors in quantum integrable models
solved by the algebraic Bethe ansatz \cite{Slavnov} (see also
\cite{Claeys,TF}, where similar relations are applied to the
Gaudin models).
 The proof of the determinant identities uses
factorization formulae for the Lax matrices \cite{VZ} which take
their origin as classical limits (and rational degenerations) of
expressions for the factorized $L$-operators through intertwining
vectors participating in the so-called vertex-IRF correspondence
\cite{Hasegawa,Quano}. Imposing the Bethe equations in the elements
of the matrix $\widetilde {\mathcal{L}}$ and using the identity
(\ref{w15}) again, one arrives at $\det \Bigl ({\mathcal{L}}-\la
I\Bigr )=(\omega -\la )^{N-M}(-\omega -\lambda )^M$ which is
(\ref{w11}).

In this paper we suggest
an extension of (\ref{w13})--(\ref{w15}) for other root systems (see (\ref{w50}) and
(\ref{w53}) below).

Let us remark that
 a natural way to understand the quantum-classical duality
 and to explain the identification (\ref{w10}) is to
 consider a ``non-stationary version'' of the Gaudin spectral problems which is
 the set of
the Knizhnik-Zamolodchikov equations
 \beq\label{w12}
 \begin{array}{c}
 \displaystyle{
 \kappa \p_{z_i}\Psi={\bf H}_i^{\hbox{\tiny{G}}}
 \Psi\,,\quad\Psi\in{\mathcal H}\,,\quad i=1\,, \ldots \,,N.
 }
 \end{array}
  \eq
In the stationary (``quasiclassical'') limit $\kappa \rightarrow 0$ the solutions
to (\ref{w12}) admit the
expansion \cite{RV}: $\Psi=(\Psi_0+\kappa \Psi_1+ \ldots )e^{S/\kappa }$ with some function
$S=S(z_1, \ldots , z_N)$. Then,
in the leading order in $\kappa$, we get the spectral problems
(\ref{w03}) with $\psi=\Psi_0$ and
$H_i^{\hbox{\tiny{G}}}=\p_{z_i}S$. The explicit constructions of the
Matsuo-Cherednik
 type\footnote{The Matsuo-Cherednik construction \cite{MC} is a projection
 of solutions to the Knizhnik-Zamolodchikov equations (\ref{w12})
 onto some special vector
 $\langle\Omega|$ in such a way that $\langle\Omega|\Psi\rangle$ become eigenfunctions
 of the quantized version of the many-body Hamiltonian (\ref{w07}) of the Calogero-Moser type.}
  reproducing the quantum-classical duality in the
stationary limit are discussed in \cite{FV}.
%
%

{\bf Purpose of the paper} is as follows. It is widely known that
the Calogero-Moser model (\ref{w07})--(\ref{w09}) is related to
the root system $A_{N-1}$ and that it can be extended to other root systems
of simple Lie algebras \cite{OP}. In this paper we show that the
Calogero-Moser models associated
with the other classical root systems, i.e. with the root systems of BCD
types, are quantum-classically dual to the
quantum Gaudin magnets with boundary.
Note that magnets of this type previously appeared
in the context of the Knizhnik-Zamolodchikov equations
\cite{Cherednik}. The
Matsuo-Cherednik construction provides a relation between the quantum
Calogero-Moser models of BCD types and the boundary
Knizhnik-Zamolodchikov equations. In this respect the relation
between the classical Calogero-Moser models of BCD types and the
quantum Gaudin magnets with boundary could be anticipated.
Our aim is to describe
it explicitly using the algebraic Bethe ansatz method and
determinant identities of the type (\ref{w15}) for the classical Lax
matrices.

Let us describe the structure of the paper.
The Calogero-Moser models for the classical root systems B, C and D
are discussed in Section 2. Next, in Section 3, we
proceed to the quantum Gaudin magnets with boundary. We present their
solutions following \cite{Hikami}. The new result, the duality between
the quantum Gaudin magnets with boundary and the classical
Calogero-Moser models associated with the root systems B, C and D,
is the subject of Section 4. This result is established by means
of nontrivial determinant identities for characteristic polynomials of
the Lax matrices of the Calogero-Moser models which generalize
(\ref{w15}) to the other root systems. A sketch of proof of the determinant
identities is given in the appendix. Section 5 contains concluding remarks
and directions for further study.


\section{Calogero-Moser models for classical root systems}
 \setcounter{section}{2}
 \setcounter{equation}{0}
Following \cite{OP}, let us consider an extension of the Calogero-Moser
model with the Hamiltonian
  \begin{equation}
 \label{w20}
 H = \frac{1}{2} \sum_{a=1}^{N} p_a^2 - g_2^2
\sum_{a<b}^{N} \Big(\frac1{(q_a-q_b)^2} + \frac1{(q_a+q_b)^2}\Big) -
g_4^2\sum_{a=1}^{N} \frac1{(2q_a)^2} - g_1^2 \sum_{a=1}^{N}
\frac1{q_a^2}
 \end{equation}
depending on three coupling constants $g_2$, $g_4$ and $g_1$. Notice
that in the rational case the parameters $g_1$ and $g_4$ can be
unified so that the Hamiltonian (\ref{w20}) essentially depends on two
continuous parameters rather than three. However, the Hamiltonian (\ref{w20})
with three coupling constants naturally appears as the rational limit of its
trigonometric version in which all the three parameters are independent.

 The Lax pair for the model (\ref{w20})
 is the following pair of $(2N+1)\times(2N+1)$ block-matrices:
 \begin{equation}\label{w21}
L =\left(
    \begin{array}{ccc}
    P+A & B & C     \\
    -B & -P-A & -C    \\
    -C^T & C^T & 0
        \end{array}
    \right),
 \quad\quad\quad
 M =\left(
    \begin{array}{ccc}
    \check{A}+d & \check{B} & \check{C}    \\
    -\check{B} & -\check{A}+d & -\check{C} \\
    -\check{C}^T & \check{C}^T & d_0 \\
        \end{array}
     \right).
 \end{equation}
Here $P,A,B$ are matrices of size $N\times N$, $C$ is a column of
length $N$ with the entries
 \beq\label{w22}
  \begin{array}{c}
  \displaystyle{P_{ab} = {\dot q}_a}\delta_{ab}\,,\quad
  \displaystyle{A_{ab} = \frac{g_2 (1-\delta_{ab})}{q_a-q_b} }\,,
  \quad(C)_{a} =\frac{g_1}{q_a}\,,\quad
  \displaystyle{B_{ab} = \frac{g_2 (1-\delta_{ab})}{q_a+q_b} +
  \frac{g_4\sqrt{2}\delta_{ab}}{2q_a}
  }
  \end{array}
 \eq
and
 \beq\label{w227}
  \begin{array}{c}
  \displaystyle{\check{A}_{ab} = -\frac{g_2 (1-\delta_{ab})}{(q_a-q_b)^2} }\,,
  \quad(\check{C})_{a} =-\frac{g_1}{q_a^2}\,,\quad
  \displaystyle{\check{B}_{ab} = -\frac{g_2 (1-\delta_{ab})}{(q_a+q_b)^2}
  -
  \frac{g_4\sqrt{2}\delta_{ab}}{2q_a^2}
  }
  \,,
  \end{array}
 \eq
 where  $a,b=1, \ldots ,N$.
In the $M$-matrix one also has the diagonal matrix
 $d_{ab}=\delta_{ab}d_a$ with
 \beq\label{w23}
  \begin{array}{c}
  \displaystyle{
  d_{a} = \frac{g_1^2}{g_2} \frac1{q_a^2} +g_4\sqrt{2} \frac1{(2q_a)^2} +g_2 \sum\limits_{c \neq a}
    \Big(\frac1{(q_a-q_c)^2} + \frac1{(q_a+q_c)^2}\Big)
  }\,,\quad
  \displaystyle{d_0 = 2g_2 \sum\limits_{c=1}^N \frac1{q_c^2}\,.
  }
  \end{array}
 \eq
The superscript $T$  denotes transposition, and the checks in the
$M$-matrix mean the derivatives of the corresponding functions
entering
 the Lax matrix.

The Lax equation $\dot L=[L,M]$ with the Lax
pair (\ref{w21})--(\ref{w23})
is equivalent to the equations of motion generated by
the Hamiltonian (\ref{w20}) $H=(1/4)\tr L^2$ if the additional
constraint for the coupling constants $g_2$, $g_4$ and $g_1$ holds
true:
 \begin{equation}
 \label{w24}
  g_1(g_1^2 - 2g_2^2 + \sqrt{2}g_2g_4) = 0\,.
 \end{equation}
The higher Hamiltonians in involution are $H_k=\frac{1}{2k}\, \tr L^k$.

The following special cases satisfying
(\ref{w24}) correspond to the models associated with the classical root systems:
 \begin{equation}
 \label{w241}
 \begin{array}{l}
\hbox{-- ${\rm B}_N$ (${\rm so}_{2N+1}$): $g_4=0$, $g_1^2=2g_2^2$,
the size of the Lax matrix is $(2N+1)\times(2N+1)$;}\hfill
\\ \ \\
\hbox{-- ${\rm C}_N$ (${\rm sp}_{2N}$): $g_1=0$, the size of the Lax
matrix is $2N\times2N$;}\hfill
\\ \ \\
\hbox{-- ${\rm D}_N$ (${\rm so}_{2N}$): $g_1=0$, $g_4=0$, the size
of the Lax matrix is $2N\times2N$.}\hfill
 \end{array}
 \end{equation}
Notice that in the ${\rm C}_N$ and ${\rm D}_N$ cases the Lax
matrix has effective size $2N\times 2N$, since in these cases we can
remove the row and the column proportional to $g_1$.
The redefinition $M\rightarrow M-d_0 1_{2N+1}$ (here $1_{2N+1}$ is the matrix
with the only nonzero element $1$ at the south-east corner) which does not spoil
the Lax equation makes
the effective size of the $M$-matrix the same ($2N\times 2N$).

Let us also mention that the Hamiltonian (\ref{w20}) can be
straightforwardly  generalized to the trigonometric (and elliptic) case
\cite{OP}, where the constants $g_1$ and $g_4$ are not unified into
a single combination $g_1^2+g_4^2/4$. The model (\ref{w20}) is
associated with the ${\rm BC}_N$ root system, and the classical root
systems (\ref{w241}) appear as its particular cases: $g_4=0$ for
${\rm B}_N$, $g_1=0$ for ${\rm C}_N$ and $g_1=g_4=0$ for  ${\rm
D}_N$.
Moreover, the model (\ref{w20}) is integrable for all arbitrary
constants without restriction (\ref{w24}), see \cite{Feher}. In this
paper we deal with the Lax representation (\ref{w21}) (which
requires (\ref{w24})) since we found the determinant identities for
these Lax matrices. This is why the case ${\rm B}_N$ requires
separate consideration (the Lax matrix is of size
$(2N+1)\times(2N+1)$), while ${\rm D}_N$ can be considered as
a particular case of ${\rm C}_N$ model with $g_4=0$.

\paragraph{Factorization formulae.}
In what follows we need the factorization formulae for the Lax
matrix (\ref{w21}). Let us briefly recall the corresponding results
from \cite{VZ}.

\noindent \underline{{Factorization for the $C_{N}$ and $D_{N}$ root
systems.}} Introduce a set of $2N\times 2N$ matrices. The first one
is the diagonal matrix
 \beq\label{w54}
\begin{array}{c}
    \displaystyle{
        D^{0}_{ij} = \delta_{ij} \left\{\begin{array}{l}
            2q_{i}\prod\limits_{k \neq i}^{N}(q_{i}-q_{k})(q_{i}+q_{k})\,,\  i \leq N\,,\\
            -2q_{i-N}\prod\limits_{k \neq i-N}^{N}(q_{i-N}-
            q_{k})(q_{i-N}+q_{k})\,,\  N+1 \leq i \leq
            2N\,,
        \end{array}\right.
    }
\end{array}
 \eq
the next one is the Vandermonde matrix
 \begin{equation}
 \label{w55}
  V_{ij} =
  \left\{ \begin{array}{l} q_{i}^{j-1},\  i \leq N\,,
  \\ \ \\
  (-q_{i-N})^{j-1},\  N+1 \leq i \leq 2N\,,
 \end{array}\right.
 \end{equation}
and finally we introduce two nilpotent matrices
 \beq\label{w56}
 \begin{array}{c}
 (C_{0})_{ij}= \left\{\begin{array}{l}
 j-1\  \hbox{if}\ j=i+1,\\ \\
 0\  \hbox{otherwise}\,,
 \end{array}\right.
 \qquad
\tilde{C}_{ij} = \left\{ \begin{array}{l} 1\  \hbox{if}\ j=i+1\
\hbox{and}\ j \ \hbox{even}\,,
 \\ \\
0\  \hbox{otherwise}\,.
 \end{array}\right.
 \end{array}
\eq
They are strictly triangular matrices with zeros on the main diagonal.
 Consider the $2N\times 2N$ matrix $L'$ obtained from the Lax matrix
 (\ref{w21}) in the ${\rm C}_N$ case (\ref{w241}) as follows.
 Set the coupling constants $g_1=0$, $g_2=\hbar$, $g_4=\sqrt{2}\hbar\xi$ and
 make the substitutions
 \begin{equation}
 \label{w562}
  \displaystyle{
  {\dot q}_i\rightarrow \frac{\xi\hbar}{q_{i}}+
  \sum\limits_{k\neq i}^{N}\Big(\frac{\hbar}{q_{i}-q_{k}}
+\frac{\hbar}{q_{i}+q_{k}}\Big)\,,\quad i=1, \ldots ,N\,.
 }
 \end{equation}
  Then the matrix $L'$ is represented in the form
 \begin{equation}\label{w58}
  \displaystyle{
 L'= \hbar (D^{0})^{-1}V( C_{0}-(1-2\xi)\tilde{C})V^{-1}D^{0}\,.
 }
 \end{equation}
 For $\xi =0$ (\ref{w58}) yields the $D_N$ case.

\noindent \underline{{Factorization for the $B_{N}$ root system.}}
Introduce the following set of $(2N+1)\times(2N+1)$ matrices:
 \beq\label{w59}
 \begin{array}{c}
    \displaystyle{
        D^{0}_{ij} = \delta_{ij} \left\{ \begin{array}{l}
            \sqrt{2}q_{i}^{2}\prod\limits_{k \neq i}^{N}(q_{i}-q_{k})
            (q_{i}+q_{k})\,,\ \: i \leq N\,,\\
            \sqrt{2}q_{i-N}^{2}\prod\limits_{k \neq i-N}^{N}(q_{i-N}-q_{k})
            (q_{i-N}+q_{k})\,,\ \: N+1 \leq i \leq 2N\,, \\
            \prod\limits_{k=1}^{N}(-q_{k}^{2})\,,\ \:i=2N+1
         \end{array}\right.
    }
 \end{array}
 \eq
 and
 \begin{equation}
\label{w60} V_{ij} =  \left\{ \begin{array}{l}
q_{i}^{j-1}\,,\  i \leq N\,,\\ \\
(-q_{i-N})^{j-1}\,,\  N+1 \leq i \leq 2N\,,\\ \\
\delta_{j,1}\,,\ i=2N+1\,.
 \end{array}\right.
 \end{equation}
 Consider the $(2N+1)\times (2N+1)$ matrix $L''$ obtained from the Lax matrix
 (\ref{w21}) in the ${\rm B}_N$ case (\ref{w241}) as follows.
 Set the coupling constants $g_1=\sqrt{2}\hbar$, $g_2=\hbar$, $g_4=0$ and
 make the substitutions
 \begin{equation}
 \label{w602}
  \displaystyle{
  {\dot q}_i\rightarrow \frac{2\hbar}{q_{i}}+
  \sum\limits_{k\neq i}^{N}\Big(\frac{\hbar}{q_{i}-q_{k}}
+\frac{\hbar}{q_{i}+q_{k}}\Big)\,,\quad i=1, \ldots ,N\,.
 }
 \end{equation}
 Then the matrix $L''$ is represented in the factorized form:
 \begin{equation}
\label{w61}
 L'' = \hbar (D^{0})^{-1}V(C_{0}+\tilde{C})V^{-1}D^{0},
 \end{equation}
where $C_0$ and $\ti{C}$ are the matrices defined in (\ref{w56}) but of the
size $(2N+1)\times(2N+1)$.

The proofs of the factorization formulae (\ref{w58}), (\ref{w61}) can be found
in Appendix B of the paper \cite{VZ}.

\section{Gaudin model with boundary}
 \setcounter{section}{3}
 \setcounter{equation}{0}

We start from the inhomogeneous open XXX spin chain with the Yang's $R$-matrix
$R^\eta_{12}(u)=1\otimes 1+(\eta/u){\bf P}_{12}$ satisfying the Yang-Baxter equation
\beq\label{w32a}
R^\eta_{12}(u_1-u_2)R^\eta_{13}(u_1)R^\eta_{23}(u_2)=
R^\eta_{23}(u_2)R^\eta_{13}(u_1)R^\eta_{12}(u_1-u_2),
\eq
where the both sides are matrices acting in the space $V_1\otimes V_2\otimes V_3
\cong (\mC ^2)^{\otimes 3}$.
According to the general theory of integrable models with boundary \cite{Skl},
one should also consider $K$-matrices $K^{\pm}$ which solve the
reflection equations
\beq \label{w32}
\begin{array}{c}
\displaystyle{
R^\eta_{12}(u_1-u_2)K^{-}_{1}(u_1)R^\eta_{12}(u_1+u_2)K^{-}_{2}(u_2)
}
\displaystyle{
=K^{-}_{2}(u_2)R^\eta_{12}(u_1+u_2)K^{-}_{1}(u_1)R^\eta_{12}(u_1-u_2),
}
\\ \ \\
\displaystyle{ R^\eta_{12}(-u_1+u_2)
(K^{+}_{1})^{T_1}(u_1)R^\eta_{12}(-u_1-u_2-2\eta)(K_{2}^{+})^{T_2}(u_2)
= }\hfill
\\ \ \\
\hfill\displaystyle{ =(K_{2}^{+})^{T_2}(u_2)
R^\eta_{12}(-u_1-u_2-2\eta)(K^{+}_{1})^{T_1}(u_1)R^\eta_{12}(-u_1+u_2)\,.
}
\end{array}
\eq
Here $K_1^{\pm}=K^{\pm}\otimes 1$, $K_2^{\pm}=1\otimes K^{\pm}$
and $T_{1,2}$ mean transpositions in the corresponding tensor
 components.
In this paper we consider diagonal $K$-matrices of the form
 \beq \label{w31}
\begin{array}{c}
 \displaystyle{
 K^{-}(u) =\left( \begin{array}{cc}
\displaystyle{1+\frac{\al \eta}{u}} & 0\\ \\
0 & \displaystyle{-1+\frac{\al \eta}{u}}
\end{array}\right),
}
 \qquad
 \displaystyle{ K^{+}(u) =\left( \begin{array}{cc}
\displaystyle{1-\frac{\be \eta}{u+\eta}} & 0\\ \\
0 & \displaystyle{-1-\frac{\be \eta}{u+\eta}}
\end{array}\right)
}
\end{array}
\eq
with arbitrary parameters $\alpha , \beta$.
 The transfer matrix of the integrable spin chain with boundaries with the Hilbert space
 of states ${\cal H}=\otimes _{i=1}^N V_i \cong (\mC ^2)^{\otimes N}$
 is given by
 \beq \label{w33}
\begin{array}{c}
 \displaystyle{
 {\bf T}(u) = \tr_{0}
 \Big(K^{+}_{0}(u)R^\eta_{01}(u-z_1)...R^\eta_{0N}(u-z_N)K^{-}_{0}
 (u)R^\eta_{0N}(u+z_{N})...R^\eta_{01}(u+z_1)\Big)\,,
 }
\end{array}
\eq
where the trace $\tr_{0}$ is taken in the auxiliary space $V_0\cong \mC ^2$.
This transfer matrix differs from the Sklyanin's one \cite{Skl} by an
inessential common scalar factor.
The parameters $\{z_k\}$ in (\ref{w33}) are inhomogeneity
 parameters. The Yang-Baxter equation together with the reflection equations
 imply that ${\bf T}(u)$ is a commutative family of operators:
 $[{\bf T}(u), {\bf T}(v)]=0$ for any $u,v$.

 The limit to the Gaudin model \cite{Hikami} is the limit  $\varepsilon\rightarrow 0$ in
 (\ref{w33}) performed after the substitution $\eta = \varepsilon \hbar$.
As one can readily check, in this limit
 \beq\label{w332}
  \displaystyle{
  {\bf T}(u) = 2+\varepsilon\hbar \ga(u)
+ \varepsilon^2\hbar^2 {{\bf T}}^{\hbox{\tiny{G}}}(u)+ O(\varepsilon^3)\,,
 }
 \eq
where $\displaystyle{\ga (u)=\sum_i \left (\frac{1}{u-z_i}+\frac{1}{u+z_i}\right )}$
is a scalar function and
 \beq\label{w333}
  \displaystyle{
  {{\bf T}}^{\hbox{\tiny{G}}}(u) =-\frac{2\alpha \beta}{u^2} + \frac{1}{\hbar}
 \sum\limits_{i=1}^{N}\Big(\frac{ {\bf H}_i^{\hbox{\tiny{G}}} }{u-z_i}-
 \frac{ { {\bf H}_i^{\hbox{\tiny{G}}} }}{u+z_i}\Big)\,,
 }
 \eq
where
\beq \label{w30}
 \displaystyle{
 \frac{1}{\hbar}\,{\bf H}_i^{\hbox{\tiny{G}}} =\frac{\xi \sigma_3^{(i)} }{z_i}+
  \sum\limits_{k \neq i}^{N}\Big(\frac{{\bf P}_{ik}}{z_i-z_k}+\frac{\sigma_3^{(i)}
{\bf P}_{ik}\sigma_3^{(i)} }{z_i+z_k}\Big)
 \,,\qquad \xi = \alpha -\beta
 }
 \eq
are Hamiltonians of the Gaudin model with boundary\footnote{For
simplicity, we use the same notation for them as for the Gaudin
Hamiltonians in the quasiperiodic case (\ref{w01}) discussed in the
introduction. This can not lead to a misunderstanding since the
latter will not appear in what follows.}. Note that the boundary
parameters $\alpha$, $\beta$ enter here in the combination $\xi
=\alpha -\beta$, so effectively there is only one boundary parameter
in the Gaudin limit instead of two. More general Gaudin models with
boundary are discussed in \cite{Amico2} and \cite{LIL}.

The commutativity of the transfer matrices implies that the Gaudin
Hamiltonians commute: $[{\bf H}_i^{\hbox{\tiny{G}}}, {\bf
H}_j^{\hbox{\tiny{G}}}]=0$. Therefore, one can find a complete set
of common solutions to the eigenvalue problems ${\bf
H}_i^{\hbox{\tiny{G}}}\psi = H_i^{\hbox{\tiny{G}}}\psi$ for the
Hamiltonians (\ref{w30}). The algebraic Bethe ansatz provides the
following solution:
\beq \label{w34}
\begin{array}{c}
 \displaystyle{
  \frac{1}{\hbar}\,{H}_i^{\hbox{\tiny{G}}}=\frac{1}{\hbar}\,
  {H}_i^{\hbox{\tiny{G}}}(\{z_k\}_N,\{\mu_\ga\}_M,\xi )
 }
 \\ \ \\
 \displaystyle{
  =\frac{\xi}{z_{i}}+\sum\limits_{k\neq i}^{N}\Big(\frac{1}{z_{i}-z_{k}}
+\frac{1}{z_{i}+z_{k}}\Big)-\sum\limits_{\ga=1}^{M}\Big(\frac{1}{z_{i}-
\mu_{\ga}}+\frac{1}{z_{i}+\mu_{\ga}}\Big).
 }
\end{array}
\eq
The solution depends on the set $\{z_k\}_N$ of $N$ inhomogeneity
 parameters and the set $\{\mu_\ga\}_M$ of $M\leq [N/2]$ Bethe roots.
 (As in the quasiperiodic case,
$M$ is equal to the number of overturned spins in the eigenstate $\psi$). The Bethe roots are
 solutions of the system of $M$ Bethe equations
\beq \label{w35}
\begin{array}{c}
 \displaystyle{
 \frac{2\xi}{\mu_{\ga}}
 +\sum\limits_{k=1}^{N}\Big(\frac{1}{\mu_{\ga}-z_{k}}+
 \frac{1}{\mu_{\ga}+z_{k}}\Big)=\frac{2}{\mu_{\ga}}+\sum\limits_{c\neq
\ga}^{M}\Big(\frac{2}{\mu_{\ga}-\mu_{c}}+\frac{2}{\mu_{\ga}+\mu_{c}}\Big)\,,\quad
 \ga=1, \ldots , M\,.
 }
\end{array}
\eq
Different solutions to this algebraic system
correspond to different eigenstates of the Gaudin Hamiltonians.

 As is shown in the next section, this model is quantum-classically
 dual to the Calogero-Moser models associated with the root
 systems ${\rm C}_N$ and ${\rm D}_N$. In the case of the ${\rm B}_N$ root
 system we consider the Gaudin model (\ref{w30}) with $N+1$ spins and
 the following conditions: $\xi =0$, $z_{N+1}=0$. Then the Gaudin Hamiltonians
 (\ref{w30}) take the form:
\beq \label{w36}
 \displaystyle{
 \frac{1}{\hbar}\,\tilde{\bf H}_i^{\hbox{\tiny{G}}} =
 \frac{{\bf P}_{i,N+1}}{z_i}+\frac{\sigma_3^{(i)}
{\bf P}_{i,N+1}\sigma_3^{(i)} }{z_i}+
  \sum\limits_{k \neq i}^{N}\Big(\frac{{\bf P}_{ik}}{z_i-z_k}+\frac{\sigma_3^{(i)}
{\bf P}_{ik}\sigma_3^{(i)} }{z_i+z_k}\Big)\,,\quad i=1, \ldots ,N\,.
 }
 \eq
In the case $z_{N+1}=0$ the Hamiltonian ${\bf H}_{N+1}^{\hbox{\tiny{G}}}$
 disappears from the transfer matrix.
For the eigenvalues of the Hamiltonians (\ref{w36}) we have:
 \beq \label{w37}
\begin{array}{c}
 \displaystyle{
  \frac{1}{\hbar}\,\tilde{H}_i^{\hbox{\tiny{G}}}(\{z_k\}_N,
  \{\mu_\ga\}_M)=\frac{2}{z_{i}}+\sum\limits_{k\neq
i}^{N}\Big(\frac{1}{z_{i}-z_{k}}
+\frac{1}{z_{i}+z_{k}}\Big)-\sum\limits_{\ga=1}^{M}\Big(\frac{1}{z_{i}-
\mu_{\ga}}+\frac{1}{z_{i}+\mu_{\ga}}\Big)\,,
 }
\end{array}
\eq
 and the Bethe equations are of the form
 \beq \label{w38}
\begin{array}{c}
 \displaystyle{
\sum\limits_{k=1}^{N}\Big(\frac{1}{\mu_{\ga}-z_{k}}+\frac{1}{\mu_{\ga}+
z_{k}}\Big)=2\sum\limits_{c\neq
\ga}^{M}\Big(\frac{1}{\mu_{\ga}-\mu_{c}}+\frac{1}{\mu_{\ga}+\mu_{c}}\Big)\,,\quad
\ga=1, \ldots ,M\,.
 }
\end{array}
\eq
%


\section{Statement of duality}
 \setcounter{section}{4}
 \setcounter{equation}{0}

Now we are ready to formulate the main result of the paper.

\noindent{\bf Theorem.} {\em  Let us identify the marked
points $z_i$ in the Gaudin model with coordinates of the Calogero-Moser
particles
 \beq\label{w40}
    \displaystyle{
 z_j=q_j\,,\quad j=1\,, \ldots \,,N
 }
 \eq
 and make the substitution
 \beq\label{w41}
    \displaystyle{
\dot q_j= H_j^{\hbox{\tiny{G}}}\quad\hbox{or}\quad \dot q_j=
{\tilde H}_j^{\hbox{\tiny{G}}}\,,
 \ \ j=1\,, \ldots \,,N
 }
 \eq
 in the Lax matrix (\ref{w21}), which we denote as $L({\{\dot q_j \}}, {\{q_j \}}
 |\,g_1,g_2,g_4)$. Here $H_j^{\hbox{\tiny{G}}}$ and ${\tilde H}_j^{\hbox{\tiny{G}}}$
 are eigenvalues of the Gaudin Hamiltonians (\ref{w34}) and (\ref{w37}) respectively.
 For the classical root systems (\ref{w241})
 set the coupling constants as follows:
 $$
 \begin{array}{l}
\hbox{${\rm B}_N:$ ${\tilde H}_j^{\hbox{\tiny{G}}}$ from
(\ref{w37}), $g_1=\sqrt{2}\hbar$, $g_2=\hbar$, $g_4=0$,
 the Lax matrix size is
$(2N+1)\times(2N+1)$;}
\\ \ \\
\hbox{${\rm C}_N:$  $H_j^{\hbox{\tiny{G}}}$ from (\ref{w34}),
$g_1=0$, $g_2=\hbar$, $g_4=\sqrt{2}\hbar \, \xi$, the Lax matrix size
is $2N\times2N$;}
\\ \ \\
\hbox{${\rm D}_N:$ $H_j^{\hbox{\tiny{G}}}$ from (\ref{w34}) with
$\xi =0$, $g_1=0$, $g_2=\hbar$, $g_4=0$,  the Lax matrix size is
$2N\times2N$.}
 \end{array}
 $$
If, for any $M \leq [N/2]$, the Bethe roots $\{\mu_\ga\}$ satisfy the Bethe
equations (more precisely, (\ref{w38}) for the ${\rm B}_N$ case,
(\ref{w35}) for the ${\rm C}_N$ case and  (\ref{w35}) with $\xi =0$ for
the ${\rm D}_N$ case), i.e.,
$H_j^{\hbox{\tiny{G}}}$ or ${\tilde H}_j^{\hbox{\tiny{G}}}$ belong to the spectrum
of the Gaudin model,
then all eigenvalues of the Lax matrix
 $L({\{{\tilde H}_j^{\hbox{\tiny{G}}}\}}, {\{q_j \}}|\,g_1,g_2,g_4)$ or
 $L({\{H_j^{\hbox{\tiny{G}}}\}}, {\{q_j \}}|\,g_1,g_2,g_4)$ and,
 therefore, all the integrals of motion, are equal to zero.
}

This result means that
the spectrum of the Hamiltonians of the quantum Gaudin magnets
with boundary is reproduced from the intersection of two Lagrangian manifolds
in the phase space of the $N$-body
Calogero-Moser models associated with the classical root systems.
One of these Lagrangian manifolds is the hyperplane
corresponding to fixing all coordinates of the particles
and the other one is the level zero set of $N$ higher classical Hamiltonians in involution, i.e.,
the $N$-dimensional hypersurface obtained by putting
all the integrals of motion equal to zero. This result looks similarly to the
particular case for the ${\rm A}_{N-1}$ root
system (the case when the twist matrix in the Gaudin model is absent)
discussed in \cite{MTV}.

The proof of the quantum-classical duality between the Calogero-Moser models of
the type (\ref{w20}) and the Gaudin models (\ref{w30}), (\ref{w36}) is
based on the nontrivial determinant identities which are discussed below.

\paragraph{The determinant identities.} The statement of the theorem is equivalent to the following
relations which are valid assuming that
the parameters $\{\mu_{\gamma}\}$ satisfy the Bethe equations
(i.e., that the Bethe vectors $\psi$ are ``on-shell''), so that
$H_j^{\hbox{\tiny{G}}}$ and ${\tilde H}_j^{\hbox{\tiny{G}}}$ do belong to the
spectrum of the Gaudin models:
 \beq
 \label{w42}
 \displaystyle{
 \det\limits_{(2N+1)\times(2N+1)}\Big[L\Big(\{\tilde{H}_j^{\hbox{\tiny{G}}}\}, \{q_j\}|\,
 \sqrt{2}\hbar,\hbar,0\Big)-\la I\Big]
  =-\la^{2N+1}
  }
 \eq
 for the ${\rm B}_N$ case and
 \beq
 \label{w43}
 \displaystyle{
 \det\limits_{2N\times2N}\Big[L\Big(\{{H}_j^{\hbox{\tiny{G}}}\}, \{q_j\}|\,0,
 \hbar,\sqrt{2}\hbar \, \xi \Big)-\la I\Big]
  =\la^{2N}
  }
 \eq
for the ${\rm C}_N$ case. The ${\rm D}_N$ case is obtained from (\ref{w43})
when $\xi =0$ ($\al=\be$). In
(\ref{w42}) $\tilde{H}_j^{G}$ is the set
$\tilde{H}_j^{G}(\{q\}_N,\{\mu\}_M)$ from (\ref{w37}), and in
(\ref{w43}) ${H}_j^{G}=\tilde{H}_j^{G}(\{q\}_N,\{\mu\}_M,\xi )$
from (\ref{w34}).

In order to prove (\ref{w42})--(\ref{w43}) we need certain determinant identities.

\noindent \underline{{The identity for the ${\rm B}_N$ case}}.
Let us introduce the
notation
 \beq
 \label{w45}
 \displaystyle{
 {\mathcal
 L}=L\Big(\{\tilde{H}_j^{\hbox{\tiny{G}}}\}_N,\{q_j \}_N|\,\sqrt{2}\hbar,\hbar,0\Big)\,.
  }
 \eq
This is just the Lax matrix (\ref{w21}) of the ${\rm B}_N$ type and of
size $(2N+1)\times(2N+1)$ in which the set
$\{\tilde{H}_j^{\hbox{\tiny{G}}}\}_N$ is taken from (\ref{w37}). Along with
$\{q_j\}$, it
depends on $M$ variables $\{\mu_\ga\}$ which are regarded here as free
parameters (i.e., they are not supposed to satisfy the Bethe equations). The coupling constants
$g_1=\sqrt{2}\, \hbar$, $g_2=\hbar$, $g_4=0$
are the ones
given in the argument in the r.h.s. of (\ref{w45}).
Next, introduce the $2M\times 2M$ matrix
 \beq
 \label{w46}
 \displaystyle{
 \tilde{\mathcal
 L}=L\Big(\{-{H}_a^{\hbox{\tiny{G}}}\}(\{\mu_\ga\}_M,\{q_j\}_N,\xi =-1)|\,0,
 \hbar,\sqrt{2}\hbar\Big)\,,
  }
 \eq
where the arguments $\{q\}_N$ and $\{\mu\}_M$ are interchanged in
the expression (\ref{w34}). More precisely, $\tilde{\mathcal L}$ is
the block-matrix
 \beq
 \label{w47}
 \displaystyle{
 \tilde{\mathcal L}
 =\mats{\ti A}{\ti B}{-\ti B}{-\ti A}\,,\quad  \ti A\,,\ti B\in \MatM \,,
  }
 \eq
 where for $i,j=1, \ldots ,M$
 \beq
 \label{w48}
 \displaystyle{
 {\ti A}_{ij}=\delta_{ij}\Big[\frac{\hbar}{\mu_i} +
 \sum\limits_{k=1}^{N}\Big(\frac{\hbar}{\mu_i-q_k}+\frac{\hbar}{\mu_i+q_k}\Big)
  - \sum\limits_{l \neq i}^{M}\Big(\frac{\hbar}{\mu_i-\mu_l}+
  \frac{\hbar}{\mu_i+\mu_l}\Big)\Big] + \frac{\hbar(1-\delta_{ij})}{\mu_i-\mu_j}
  }
 \eq
 and
 \beq
 \label{w49}
 \displaystyle{
 {\ti B}_{ij}=\delta_{ij}\frac{\hbar}{\mu_i}+(1-
\delta_{ij})\frac{\hbar}{\mu_i+\mu_j}\,.
  }
 \eq
Then the identity is as follows:
 \beq
 \label{w50}
 \displaystyle{
 \det\limits_{(2N+1)\times(2N+1)}\Big({\mathcal L}-\la I \Big)
  =-\la^{2N-2M+1}\det\limits_{2M\times2M}\Big( \tilde{\mathcal
  L}-\la I \Big)\,.
  }
 \eq

 \paragraph{Example.} Let us give a simple example for $N=M=1$:
 \beq \label{E1}
 \mathcal{L} = \left(\begin{array}{ccc}
 \displaystyle{\ti{H}^{\rm G}_1(\{q\}_1,\{\mu\}_1)} &  \displaystyle{0} &
 \displaystyle{\frac{\sqrt{2}\hbar}{q_1}} \\
  \displaystyle{0} &  \displaystyle{-\ti{H}^{\rm G}_1(\{q\}_1,\{\mu\}_1)} &
  \displaystyle{-\frac{\sqrt{2}\hbar}{q_1}}\\
  \displaystyle{-\frac{\sqrt{2}\hbar}{q_1}} &
  \displaystyle{\frac{\sqrt{2}\hbar}{q_1}} & \displaystyle{0}
 \end{array}\right),
 \eq
 where $\displaystyle{\ti{H}^{\rm G}_1(\{q\}_1,\{\mu\}_1)=
 \frac{2\hbar}{q_1}-\frac{\hbar}{q_1-\mu_1}-\frac{\hbar}{q_1+\mu_1}}$,
 \beq \label{E2}
 \ti{\mathcal{L}} = \left( \begin{array}{cc}
 \displaystyle{-H^{\rm G}_1(\{\mu\}_1,\{q\}_1,-1)} & \displaystyle{\frac{\hbar}{\mu_1}}\\
 \displaystyle{-\frac{\hbar}{\mu_1}} & \displaystyle{H^{\rm G}_1(\{\mu\}_1,\{q\}_1,-1)}
 \end{array}\right),
 \eq
 where $\displaystyle{H^{\rm G}_1(\{\mu\}_1,\{q\}_1,-1)=
 -\frac{\hbar}{\mu_1}-\frac{\hbar}{\mu_1-q_1}-\frac{\hbar}{\mu_1+q_1}}$.
 One can directly calculate the characteristic polynomials of these
 matrices and find that (\ref{w50}) holds true.

\noindent \underline{{The identity for the ${\rm C}_N$ and ${\rm D}_N$
cases}}. Similarly, let us introduce the $2N\times 2N$ matrix
 \beq
 \label{w51}
 \displaystyle{
 {\mathcal
 L}=L\Big(\{{H}_j^{\hbox{\tiny{G}}}\}_N(\{q\}_N,\{\mu\}_M,\xi ),
 \{q\}_N|\,0,\hbar,\sqrt{2}\, \hbar \, \xi \Big)\,,
  }
 \eq
 which is the matrix (\ref{w21}) of the type ${\rm C}_N$ with the set
 $\{{H}_j^{\hbox{\tiny{G}}}\}_N=\{{H}_j^{\hbox{\tiny{G}}}\}_N(\{q\}_N,\{\mu\}_M,\xi )$ from
 (\ref{w34}) depending on the $M$ independent variables $\{\mu_\ga\}_M$.
The dual matrix $\tilde {\cal L}$ has size $2M\times 2M$:
 \beq
 \label{w52}
 \displaystyle{
 \ti{\mathcal
 L}=L\Big(\{{H}_a^{\hbox{\tiny{G}}}\}_M(\{\mu\}_M,\{q\}_N,1-
 \xi ),\{\mu \}_M|\,0,\hbar,\sqrt{2}\, \hbar \, (1-\xi )\Big)\,.
  }
 \eq
The arguments of
 the eigenvalues of the Gaudin Hamiltonians are interchanged and the
 coupling constants are different.
For the pair of matrices (\ref{w51}), (\ref{w52}) the
 identity is
 \beq
 \label{w53}
 \displaystyle{
 \det\limits_{2N\times2N}\Big({\mathcal L}-\la I\Big)
  =\la^{2N-2M}\det\limits_{2M\times2M}\Big( \tilde{\mathcal
  L}-\la I\Big)\,.
  }
 \eq

 \paragraph{Example.} Let us give a simple example
for $N=2$, $M=1$. We have:
 \begin{equation}\label{E3}
  \begin{array}{c}
 \displaystyle{
 \mathcal{L} =
 }
  \end{array}
 \end{equation}
 $$
 \displaystyle{
 =\left(  \begin{array}{cccc}
 \displaystyle{
 H^{\rm G}_1(\{q\}_2,\{\mu\}_1,\xi)} & \displaystyle{ \frac{\hbar}{q_1-q_2}}
 & \displaystyle{\frac{\hbar \xi}{q_1}} & \displaystyle{\frac{\hbar}{q_1+q_2}
 }
 \\ \ \\
 \displaystyle{
 \frac{\hbar}{q_2-q_1}} & \displaystyle{H^{\rm G}_2(\{q\}_2,\{\mu\}_1,\xi)}
 & \displaystyle{\frac{\hbar}{q_1+q_2}} & \displaystyle{\frac{\hbar \xi}{q_2}
 }
 \\ \ \\
 \displaystyle{
 -\frac{\hbar \xi}{q_1}} & \displaystyle{-\frac{\hbar}{q_1+q_2}}
 & \displaystyle{-H^{\rm G}_1(\{q\}_2,\{\mu\}_1,\xi)} & \displaystyle{\frac{\hbar}{q_2-q_1}
 }
 \\ \ \\
 \displaystyle{
 -\frac{\hbar}{q_1+q_2}} & \displaystyle{-\frac{\hbar \xi}{q_2}}
 & \displaystyle{\frac{\hbar}{q_1-q_2}} & \displaystyle{-H^{\rm G}_2(\{q\}_2,\{\mu\}_1,\xi)
 }
 \end{array} \right)
 }
 $$
 where
 $$
 H^{\rm G}_1(\{q\}_2,\{\mu\}_1,\xi)=\frac{\xi \hbar}{q_1}
 +\frac{\hbar}{q_1-q_2}+\frac{\hbar}{q_1+q_2}-\frac{\hbar}{q_1-\mu_1}-\frac{\hbar}{q_1+\mu_1},
 $$
 $$
 H^{\rm G}_2(\{q\}_2,\{\mu\}_1,\xi)=\frac{\xi \hbar}{q_2}
 +\frac{\hbar}{q_2-q_1}+\frac{\hbar}{q_2+q_1}-\frac{\hbar}{q_2-\mu_1}-\frac{\hbar}{q_2+\mu_1},
 $$
and
 \beq \label{E4}
 \ti{\mathcal{L}} = \left(\begin{array}{cc}
 \displaystyle{H^{ \rm G}_1 (\{\mu\}_1,\{q\}_2,1-\xi)} &
 \displaystyle{\frac{\hbar (1-\xi)}{\mu_1}} \\
 \displaystyle{-\frac{\hbar (1-\xi)}{\mu_1}} &
 \displaystyle{-H^{ \rm G}_1 (\{\mu\}_1,\{q\}_2,1-\xi)}
 \end{array}\right),
 \eq
 where
 $$
H^{ \rm G}_1 (\{\mu\}_1,\{q\}_2,1-\xi)=\frac{(1-\xi )\hbar}{\mu_1}-
\frac{\hbar}{\mu_1-q_1}-\frac{\hbar}{\mu_1-q_2}-\frac{\hbar}{\mu_1+q_1}-
\frac{\hbar}{\mu_1+q_2}\,.
$$
 One can directly verify that these matrices satisfy (\ref{w53}).

The proof of the identities (\ref{w50}) and (\ref{w53}) requires
cumbersome calculations. In the $M=0$ or $N=0$ cases they follow from the
factorization formulae for the Lax matrices (\ref{w58}), (\ref{w61}) \cite{VZ}. In
the general case the proof also uses the factorization formulae. It
is similar to the one presented in \cite{GZZ} for (\ref{w15}) but
 the calculations are more complicated. We give a sketch of proof of (\ref{w53}) in the
 appendix. The identity (\ref{w50}) is proved in a similar way.
\paragraph{Factorization formulae.}
 In the simplest cases when the set of the parameters $\{ \mu \}$ is empty ($M=0$) the
 determinant identities directly follow from the factorization formulae.
 Indeed, consider the ${\rm C}_{N}$ and ${\rm D}_{N}$ root
systems. Then, for  $M=0$, the matrix (\ref{w51}) admits the
 factorization (\ref{w58}):
 \begin{equation}
 \label{w058}
 \begin{array}{c}
\displaystyle{
  \left. \phantom{\int}{\cal L} \right |_{M=0}\!=
  L\Big(\{{H}_j^{\hbox{\tiny{G}}}\}_N(\{q\}_N,\{ \mu \}_0,\xi),
 \{q\}_N|\,0,\hbar,\sqrt{2}\, \hbar \, \xi \Big)=
 }
 \\ \ \\
\displaystyle{
 =
\hbar\, (D^{0})^{-1}V( C_{0}-(1-2\xi)\tilde{C})V^{-1}D^{0}\,,
 }
 \end{array}
 \end{equation}
 where $\{\mu \}_0$ means that this set of parameters is empty.
At $\xi =0$ (\ref{w058}) reproduces the matrix for the ${\rm D}_N$ case.
Similarly, for  the ${\rm B}_{N}$ root system the matrix (\ref{w45}) is
factorized as in (\ref{w61}). It can be rewritten as follows:
 \begin{equation}
\label{w061} \left. \phantom{\int}{\cal L} \right |_{M=0}\!=
L\Big(\{{\ti{H}}_j^{\hbox{\tiny{G}}}\}_N(\{q\}_N,\{\mu \}_0),
 \{q\}_N|\,\sqrt{2}\hbar,\hbar,0 \Big)\, =
\hbar (D^{0})^{-1}V(C_{0}+\tilde{C})V^{-1}D^{0},
 \end{equation}
where $C_0$ and $\ti{C}$ are taken from (\ref{w56}) but now they are of size
$(2N+1)\times(2N+1)$.
In both cases (\ref{w058}) and (\ref{w061}) the matrices are gauge
equivalent (i.e. conjugated) to certain linear combinations of $C_0$ and
$\tilde{C}$ which are nilpotent matrices (strictly triangular matrices
with zeros on the main diagonal). In this way we get the
identities (\ref{w50}) and (\ref{w53}) for $M=0$:
$$
\left. \det_{2N\times 2N}\Bigl ({\cal L}-\lambda I\Bigr )\right |_{M=0}=\lambda^{2N}
\quad \mbox{or} \quad
\left. \det_{(2N+1)\times (2N+1)}\Bigl ({\cal L}-\lambda I\Bigr )
\right |_{M=0}=-\lambda^{2N+1}.
$$

\paragraph{Proof of duality.} Having identities (\ref{w50}) and
(\ref{w53}), one can prove the statements of the theorem (\ref{w42}),
(\ref{w43}). Let us prove (\ref{w43}) corresponding to the ${\rm C}_N$
root system. Plugging all the data from the theorem to the $2N\times
2N$ Lax matrix (\ref{w21}) and using (\ref{w53}) we have:
 \beq\label{w62}
\begin{array}{c}
\displaystyle{ \det\limits_{2N\times
2N}\Big[L\Big((H^{\hbox{\tiny{G}}}(\{q\}_N,\{\mu\}_M,\xi )|\,0,
\hbar,\sqrt{2}\hbar \, \xi \Big)-\la I\Big]= }
\\ \ \\
=\displaystyle{ \la^{2N-2M}\det\limits_{2M\times
2M}\Big[L\Big((H^{\hbox{\tiny{G}}}(\{\mu\}_M,\{q\}_N,1-\xi )|\,0,\hbar,
\sqrt{2}\hbar(1-\xi )\Big)-\la I\Big]\,.
 }
\end{array}
\eq
Let us simplify the expression for the (eigenvalues of the)
Gaudin Hamiltonians entering the determinant in the r.h.s. using the Bethe
equations (\ref{w35}):
 \beq\label{w63}
\begin{array}{c}
\displaystyle{  \left.
\phantom{\int}H^{\hbox{\tiny{G}}}_i(\{\mu\}_M,\{q\}_N,1-\xi )\right
|_{BE}=}
 \\ \ \\
 \displaystyle{ =
\frac{\hbar(1-\xi )}{\mu_i}+\sum\limits_{k\neq
i}^{M}\Big(\frac{\hbar}{\mu_i-\mu_k}+\frac{\hbar}{\mu_i+\mu_k}\Big)-
\sum\limits_{l=1}^{N}\Big(\frac{\hbar}{\mu_i-q_l}+\frac{\hbar}{\mu_i+q_l}\Big)
}
\\ \ \\
\displaystyle{ =\frac{\hbar(\xi -1)}{\mu_i}-\sum\limits_{k\neq
i}^{M}\Big(\frac{\hbar}{\mu_i-\mu_k}+\frac{\hbar}{\mu_i+\mu_k}\Big)=-
H_i^{G}(\{\mu\}_M,\{q\}_0,1-\xi )\,. }
\end{array}
\eq
Therefore, we can rewrite (\ref{w62}) as
 \beq\label{w64}
  \begin{array}{c}
\displaystyle{ \la^{2N-2M}\det\limits_{2M\times 2M}\Bigl (L(-
H^{\hbox{\tiny{G}}}(\{\mu\}_M,\{q\}_0,1-\xi
),0,\hbar,\sqrt{2}\hbar(1-\xi ))-\la I\Bigr )
 }
\\ \ \\
\displaystyle{ = \la^{2N-2M}\det\limits_{2M\times 2M}\Bigl (L^{T}(-
H^{\hbox{\tiny{G}}}(\{\mu\}_M,\{q\}_0,1-\xi
),0,-\hbar,-\sqrt{2}\hbar(1-\xi ))-\la I\Bigr )
 }
\\ \ \\
\displaystyle{ =\la^{2N-2M}\det\limits_{2M\times 2M}\Bigl (L(-
H^{\hbox{\tiny{G}}}(\{\mu\}_M,\{q\}_0,1-\xi ),0,-\hbar,-\sqrt{2}\hbar(1-\xi ))-\la I\Bigr )
}
\\ \ \\
 \displaystyle{ = \la^{2N}. }
\end{array}
\eq
 The last equality is obtained using the identity
(\ref{w53}) for $M=0$. The proof for the ${\rm B}_N$ case is similar.
$\blacksquare$

\section{Conclusion}

We have shown that
 the quantum-classical duality for the BCD root systems works in much the
 same way as for the models associated with the ${\rm A}_{N-1}$ root system including
 the existence of nontrivial determinant identities.
We have proved that the rational quantum Gaudin models with boundary
are dual to the classical rational Calogero-Moser models
 associated with the root systems of BCD types. However, an important
 difference with the ${\rm A}$-case is the absence of the twist matrix in the quantum model.

 We understand that the results obtained in this paper
 need further generalizations in several directions, including the extension to
 higher rank Gaudin magnets
 solved by means of the nested Bethe ansatz method, the
 supersymmetric extension (see \cite{TsuboiZZ} for the ${\rm A}_{N-1}$ case),
 the trigonometric version of the duality
 \cite{BLZZ}
  and the generalization to the level of spin chains and
 related Ruijsenaars-Schneider-van Diejen relativistic many-body
 systems.

 Let us also note that on the classical side we have used the Lax representation
 \cite{OP} in which the coupling constants are not arbitrary (they are assumed
 to  satisfy condition (\ref{w24})). An alternative Lax representation in matrices
 of size $2N\times 2N$ is known \cite{Feher}, which do not require
 any additional constraints. It would be interesting to connect the spectrum
 of the Lax matrix from this representation to quantum integrable models.

We hope to study the extensions and generalizations mentioned above
in future publications.


 \section*{Appendix: proof of the determinant identities}
 \def\theequation{A.\arabic{equation}}
\setcounter{equation}{0}
 Here we give a proof for the determinant identity (\ref{w53}). The
 identity (\ref{w50}) is proved in a similar way.
 The proof of (\ref{w53}) is by induction in the number $M$ of the parameters $\{ \mu \}$.
The main idea is to compare the structure of poles and
 residues of both parts of (\ref{w53}) in all the variables $\{q_j\}$
 and $\{\mu_c\}$. Before we proceed further let us formulate the following
 useful lemma.

 \noindent{\bf Lemma.} {\em The l.h.s. of (\ref{w53})
 does not have singularities at $q_i =\pm q_k$, and the r.h.s.
 does not have singularities at $\mu_a = \pm \mu_b$.}

 The proof of the lemma is identical to the one given in \cite{GZZ} for the ${\rm A}_{N-1}$
 case. One should use the factorization formulae for the Lax matrices
 from \cite{VZ}.

 Introduce the following notation:
 \beq\label{A1}
 \begin{array}{c}
 \displaystyle{
 {\mathcal L}_{N}^{M} = L\Big(\{{H}_j^{\hbox{\tiny{G}}}\}_N(\{q\}_N,\{\mu\}_M,\xi ),
 \{q\}_N|\,0,\hbar,\sqrt{2}\hbar \, \xi \Big)\,,
 }
 \\ \ \\
 \displaystyle{
  \ti{\mathcal L}_{M}^{N} = L\Big(\{{H}_a^{\hbox{\tiny{G}}}\}_M(\{\mu\}_M,
  \{q\}_N,1-\xi ),\{\mu \}_M|\,0,\hbar,
  \sqrt{2}\hbar(1-\xi )\Big)\,.
 }
 \end{array}
 \eq
 The lower index in ${\mathcal L}_{N}^{M}$ is the half of its size, and the
 upper one is the number of variables $\mu_c$ entering this matrix. Similarly, the lower
 index in $\ti{\mathcal L}_{M}^{N}$ is the half of its size, and the
 upper one is the number of the variables $q_k$.

 We prove the determinant identity by induction in the
 number $M$ of the parameters $\{ \mu \}$. For $M=0$ it holds true due to the
 factorization formula (\ref{w058}). Suppose that (\ref{w53}) is true for the
 number of the parameters $\{ \mu \}$
 equal to $M-1$. Let us study the structure of poles
 of the characteristic polynomials of the matrices (\ref{A1})
 from both sides of (\ref{w53}) in the
 first parameter $\mu_1$
 (since the determinants are symmetric functions of $\{\mu_c\}_M$ it is enough to
 consider them as functions of $\mu_1$). It is easy to see that the
 determinants have the second order poles in $\mu_1$ at
 most. Therefore, we can write:
 \beq\label{A2}
 \begin{array}{c}
 \displaystyle{
\det\limits_{2N\times 2N}\Big(\mathcal{L}^{M}_{N}-\la I\Big) = a_0
+\sum\limits_{k=1}^{N}\Big(\frac{a_k^-}{\mu_1-q_k}+\frac{a_k^+}{\mu_1+q_k}\Big)
+\sum\limits_{k=1}^{N}\Big(\frac{c_k^-}{(\mu_1-q_k)^2}+\frac{c_k^+}{(\mu_1+q_k)^2}\Big)\,,
 }
\\ \ \\
\displaystyle{ \det\limits_{2M \times 2M}
\Big(\ti{\mathcal{L}}^{N}_{M}-\la I\Big) =\ti{a}_0
+\sum\limits_{k=1}^{N}\Big(\frac{\ti{a}_k^-}{\mu_1-q_k}+\frac{\ti{a}_k^+}{\mu_1+q_k}\Big)
+\sum\limits_{k=1}^{N}\Big(\frac{\ti{c}_k^-}{(\mu_1-q_k)^2}+\frac{\ti{c}_k^+}{(\mu_1+q_k)^2}\Big)\,.
}
\end{array}
\eq
 Let us calculate all the coefficients in these expansions.
 For $a_0$ and $\tilde a_0$ we have:
 \beq\label{A3}
 \begin{array}{c}
 \displaystyle{
  a_0 = \lim\limits_{\mu_1 \rightarrow \infty}
\Big(\det\limits_{2N\times 2N}\Big(\mathcal{L}^{M}_{N}-\la I\Big)\Big)
 = \det\limits_{2N\times 2N}\Big(\mathcal{L}^{M-1}_{N} -\la I\Big)
 }
 \\ \ \\
 \displaystyle{
=\la^{2N-2M+2}
 \det\limits_{(2M-2)\times(2M-2)}\Big(\ti{\mathcal{L}}^{N}_{M-1} -\la I\Big),
 }
 \\ \ \\
 \displaystyle{
 \ti{a}_0 = \lim\limits_{\mu_1 \rightarrow \infty}
\Big(\det\limits_{2M\times
2M}\Big(\ti{\mathcal{L}}^{N}_{M}-\la I\Big)\Big) =
 \la^{2} \det\limits_{(2M-2)\times
 (2M-2)}\Big(\ti{\mathcal{L}}^{N}_{M-1}-\la I\Big)\,,
}
\end{array}
\eq
 which yields $a_0 = \la^{2N-2M}\ti{a}_0$. This is exactly what we need.

 For the coefficients in front of the second order poles
 we have:
 \beq\label{A4}
 \begin{array}{c}
 \displaystyle{
 c_k^- = \lim\limits_{\mu_1 \rightarrow q_k}\Big((\mu_1-q_k)^2
 \det\limits_{2N\times 2N}\Big(\mathcal{L}^{M}_{N}-\la I\Big)\Big)
  = -\hbar^2\det\limits_{(2N-2)\times (2N-2)}\Big(\mathcal{L}^{M-1}_{N-1}-\la I\Big)
 }
 \\ \ \\
 \displaystyle{
 =-\hbar^2\la^{2N-2M}\det\limits_{(2M-2)\times (2M-2)}\Big(\ti{\mathcal{L}}^{N-1}_{M-1}-\la I\Big),
 }
 \\ \ \\
 \displaystyle{
 \tilde{c}_k^- = \lim\limits_{\mu_1 \rightarrow q_k}\Big((\mu_1-q_k)^2
 \det\limits_{2M\times 2M}\Big(\ti{\mathcal{L}}^{N}_{M}-\la I\Big)\Big) =
  -\hbar^2\det\limits_{(2M-2)\times (2M-2)}\Big(\ti{\mathcal{L}}^{N-1}_{M-1}-\la I\Big),
 }
 \end{array}
 \eq
 which gives $c_k^- = \la^{2N-2M}\ti{c}_k^-$. The proof of $c_k^+ =
 \la^{2N-2M}\ti{c}_k^+$ is performed in the same way.

Calculation of the
  coefficients $a^{\pm}_k$ and $\ti{a}^{\pm}_k$ is more complicated.
  They are residues of the determinants in the l.h.s.
   of (\ref{A2}). Consider the pole at $\mu_1=q_1$.
   It follows from the explicit form of the matrices from (\ref{A2})
   that the residue at $\mu_1=q_1$ may come from the
   elements proportional to $\displaystyle{\frac{1}{\mu_1-q_1}}$, which is
   present in $\mathcal{L}_{11}$ and $\mathcal{L}_{N+1,N+1}$
   (and for the matrix $\ti{\mathcal{L}}$
   such elements are in $\ti{\mathcal{L}}_{11}$ and $\ti{\mathcal{L}}_{M+1,M+1}$).
   Therefore, we have linear and quadratic
   terms in $\displaystyle{\frac{1}{\mu_1-q_1}}$, and our purpose is to compute the residues.
   First, calculate the residues
   coming from the linear parts. The residue coming from  $\mathcal{L}^{M}_{N}$ is equal to
   %
   %
 \beq \label{A5}
 \begin{array}{c}
 \displaystyle{
 G=\hbar\det\limits_{(2N-1)\times(2N-1)}\left(\begin{array}{ccc}
 A-\la I& X & B\\
 -X^T& -F-\la I & -Y^T\\
 -B & Y & -A-\la I
 \end{array}\right)
 }
 \\ \ \\
 \displaystyle{
 - \hbar \det\limits_{(2N-1)\times(2N-1)}\left(\begin{array}{ccc}
 F-\la I & Y^T & X^T\\
 -Y & A-\la I & B\\
 -X & -B & -A-\la I
 \end{array}\right),
 }
 \end{array}
 \eq
 where $A$ and $B$ are the following matrices of size
 $(N-1)\times(N-1)$:
 \beq\label{A6}
 \begin{array}{c}
 \displaystyle{
 A_{ij} = \delta_{ij}\left(\frac{\hbar \xi}{q_i}+\sum\limits_{k \neq 1,i}^{N}
 \Big(\frac{\hbar}{q_i-q_k}+\frac{\hbar}{q_i+q_k} \Big) -
  \sum\limits_{l\neq 1}^{M}\Big(\frac{\hbar}{q_i-\mu_l}+\frac{\hbar}{q_i+\mu_l}\Big)\right)
 }
 \\ \ \\
 \displaystyle{
+ (1-\delta_{ij})\frac{\hbar}{q_i-q_j}, \;\;\; 2 \leq i,j \leq N,
 }
 \\ \ \\
 \displaystyle{
 B_{ij} = \delta_{ij}\frac{\hbar \xi}{q_i}+(1-\delta_{ij})
 \frac{\hbar}{q_i+q_j},\;\;\; 2\leq i,j \leq
 N\,,
 }
 \end{array}
 \eq
 while $X$ and $Y$ are columns of size $N-1$:
 \beq\label{A7}
 \begin{array}{c}
 \displaystyle{
 X_{i} = \frac{\hbar}{q_1+q_i}, \;\;\; 2 \leq i \leq N\,,
 }
 \qquad
 \displaystyle{
 Y_i = \frac{\hbar}{q_1-q_i}, \;\;\; 2\leq i \leq N\,.
 }
 \end{array}
 \eq
 Finally,  $F$ is the number
 \beq\label{A8}
 \begin{array}{c}
 \displaystyle{
F =\frac{\hbar(\xi -\frac{1}{2})}{q_1}+\sum\limits_{k \neq 1}^{N}
 \left(\frac{\hbar}{q_1-q_k}+\frac{\hbar}{q_1-q_k} \right)
 -\sum\limits_{l \neq
1}^{M}\left(\frac{\hbar}{q_1-\mu_l}+\frac{\hbar}{q_1+\mu_l}\right).
 }
 \end{array}
 \eq
 Let us split (\ref{A5}) into parts proportional to the
 expressions containing $F$ and the other ones:
 \beq \label{A91}
 \begin{array}{c}
 \displaystyle{
 G=\hbar(-F-\la)\det\limits_{(2N-2)\times(2N-2)} \left(\begin{array}{cc}
A-\la I& B\\
-B & -A-\la I
\end{array}\right)
}
\\ \ \\
\displaystyle{ - \, \hbar(F-\la)\det\limits_{(2N-2)\times(2N-2)}
\left(\begin{array}{cc}
A-\la I & B\\
-B & -A-\la I
\end{array}\right)
 }
   \end{array}
 \eq
 $$
 \displaystyle{
 +\hbar\det\limits_{(2N-1)\times(2N-1)}\left(\begin{array}{ccc}
 A-\la I & X & B\\
 -X^T& 0 & -Y^T\\
 -B & Y & -A-\la I
 \end{array}\right)
 - \hbar\det\limits_{(2N-1)\times(2N-1)}\left(\begin{array}{ccc}
0 & Y^T & X^T\\
 -Y & A-\la I & B\\
 -X & -B & -A-\la I
 \end{array}\right)\,.
 }
$$
 Therefore,
 \beq
 \label{A9}
 \begin{array}{c}
  \displaystyle{
 G=-2\hbar F\det\limits_{(2N-2)\times(2N-2)}
 \Big(\mathcal{L}^{M-1}_{N-1}-\la I\Big)+\hbar
 \det\limits_{(2N-1)\times(2N-1)}\left(\begin{array}{ccc}
 A-\la I & X & B\\
 -X^T& 0 & -Y^T\\
 -B & Y & -A-\la I
 \end{array}\right)
 }
 \\ \ \\
 \displaystyle{
 -\hbar \det\limits_{(2N-1)\times(2N-1)}\left(\begin{array}{ccc}
0 & Y^T & X^T\\
 -Y & A-\la I & B\\
 -X & -B & -A-\la I
 \end{array}\right).
 }
 \end{array}
 \eq
This is a part of the residue of $\det\limits_{2N\times 2N}\Big(\mathcal{L}^M_N-\la I\Big)$.
To compute the full residue,
we need to add the contribution coming from the second order pole at $\mu_1=q_1$.
Thereby, the full residue of $\det\limits_{2N\times 2N}
\Big(\mathcal{L}^M_N-\la I\Big)$ at $\mu_1=q_1$ is given by
\beq \label{A10}\begin{array}{c}
\displaystyle{
a^{-}_1 = -2\hbar F\det\limits_{(2N-2)\times(2N-2)}\Big(\mathcal{L}^{M-1}_{N-1}-\la I
\Big)-\hbar^2 \frac{\partial}{\partial \mu_1}\left.
 \left[ \det\limits_{(2N-2)\times(2N-2)}\left(\mathcal{L}^{M}_{N-1}-\la I\right)
 \right]\right|_{\mu_1=q_1}
}
\\ \ \\
\displaystyle{
-\hbar \det\limits_{(2N-1)\times(2N-1)}\left(\begin{array}{ccc}
0 & Y^T & X^T\\
 -Y & A-\la I & B\\
 -X & -B & -A-\la I
 \end{array}\right)
 }
 \\ \ \\
 \displaystyle{
 +\hbar \det\limits_{(2N-1)\times(2N-1)}\left(\begin{array}{ccc}
 A-\la I & X & B\\
 -X^T& 0 & -Y^T\\
 -B & Y & -A-\la I
 \end{array}\right).
}
\end{array}
\eq
For a while, let us forget about the first term in (\ref{A10}).
We claim that the remaining part vanishes:
\beq \label{A11}\begin{array}{c}
\displaystyle{
Z=\hbar \! \det\limits_{(2N-1)\times(2N-1)}\! \! \left(\begin{array}{ccc}
 A-\la I & X & B\\
 -X^T& 0 & -Y^T\\
 -B & Y & -A-\la I
 \end{array}\right)\!
 }
 \\ \ \\
 \displaystyle{
 -\! \hbar \!
 \det\limits_{(2N-1)\times(2N-1)}\!\! \left(\begin{array}{ccc}
0 & Y^T & X^T\\
 -Y & A-\la I & B\\
 -X & -B & -A-\la I
 \end{array}\right)
}
\\ \ \\
\displaystyle{
-\hbar^2 \frac{\partial}{\partial \mu_1}
\left.\left[ \det\limits_{(2N-2)\times(2N-2)}\left(\mathcal{L}^{M}_{N-1}-\la I
\right)\right]\right|_{\mu_1=q_1}=0.
}
\end{array}
\eq
This can be verified by
analyzing the structure of (\ref{A11})
as a rational function of $q_1$. As a function of $q_1$,
$Z$ has simple and second order poles at $q_1 = \pm q_k$.
Let us recall that in the first two determinants $q_1$ is contained only in the
columns $X$ and $Y$. This instantly gives us $\lim\limits_{q_1 \rightarrow \infty}Z = 0$.
Then one has to analyze the second order poles of $Z$ at $q_1 = \pm q_k$.
Notice that the last term in (\ref{A11}) contains only second order poles
and does not contain simple poles. A direct calculation shows that
the second order poles coming from the first two terms of (\ref{A11})
and the last one cancel identically.
Thus we are left with simple poles that come only from the first two terms
\beq \label{A12}
\begin{array}{c}
\displaystyle{
\hbar \det\limits_{(2N-1)\times(2N-1)}\left(\begin{array}{ccc}
 A-\la I & X & B\\
 -X^T& 0 & -Y^T\\
 -B & Y & -A-\la I
 \end{array}\right)
 }
 \\ \ \\
 \displaystyle{
 -\hbar \det\limits_{(2N-1)\times(2N-1)}\left(\begin{array}{ccc}
0 & Y^T & X^T\\
 -Y & A-\la I& B\\
 -X & -B & -A-\la I
 \end{array}\right)
 }.
 \end{array}
\eq Again, an accurate calculation provides their cancellation.
Therefore, we find that $Z=0$ and so the expression for the residue
(\ref{A10}) is significantly simplified: \beq \label{A13} a_1^- =
-2\hbar F\det\limits_{(2N-2)\times(2N-2)}
\left(\mathcal{L}_{N-1}^{M-1}-\la I\right) = -2\hbar F \,
\la^{2N-2M} \! \det\limits_{(2M-2)\times(2M-2)}
\left(\ti{\mathcal{L}}^{N-1}_{M-1}-\la I\right) , \eq where the last
equality is true due to the induction hypothesis. Similar arguments
applied to $\ti{a}_1^-$ yield \beq \label{A14} \ti{a}_1^- = -2\hbar
F
\det\limits_{(2M-2)\times(2M-2)}\left(\ti{\mathcal{L}}^{N-1}_{M-1}-\la
I\right), \eq which means that $a_1^- = \la^{2N-2M}\ti{a}_1^-$, and
so the proof of the determinant identity (\ref{w53}) is complete.

\section*{Acknowledgments}

 A. Zotov is grateful to L\'aszl\'o Feh\'er for his warm hospitality at the
 University of Szeged and the Wigner Research Centre for Physics in Budapest,
 where results of the paper were
 presented.
 The work was partially supported by RFBR grants 18-01-00273 (M.
Vasilyev, A. Zotov) and by 18-01-00461 (A. Zabrodin). The research
of A. Zabrodin was carried out within the HSE University Basic
Research Program and funded (jointly) by the  by Russian Academic
Excellence Project '5-100'. The research of A. Zotov was also
supported in part by the Young Russian Mathematics award.


\begin{small}

\end{small}

\end{document}